\documentclass[final,5p,times,twocolumn]{elsarticle}
\usepackage{amssymb}
\usepackage{lineno}
\usepackage{hyperref}
\usepackage{graphicx}
\usepackage{booktabs} 

\begin{document}
\begin{frontmatter}
\title{Resonant Ultrasound Spectroscopy of single crystalline KH$_{2}$PO$_{4}$}

\author[1]{Vikram Singh\corref{contrib}}
\author[1]{M. Bicky Singh\corref{contrib}}
\author[1]{Sunil Nair\corref{cor1}}
\ead{sunil@iiserpune.ac.in}

\address[1]{Department of Physics, Indian Institute of Science Education and Research, Dr. Homi Bhabha Road, Pune 411008, India}

\cortext[contrib]{Authors contributed equally}
\cortext[cor1]{Corresponding author}

\begin{abstract}
This study employs resonant ultrasound spectroscopy (RUS) to investigate the elastic properties of single crystalline KH$_{2}$PO$_{4}$ (KDP) through the paraelectric to ferroelectric phase transition. Noteworthy anomalies are observed in selected resonance modes and their corresponding mechanical quality factors ($Q$) around the transition temperature.The thermal evolution of elastic constants ($C_{ij}$) across the phase transition reveals a significant softening of $C_{11}$, $C_{12}$, and $C_{13}$, accompanied by a stiffening in $C_{66}$. Additionally, both $C_{33}$ and $C_{44}$ exhibit a minimum value at the transition. This anomalous behavior of all elastic constants ($C_{ij}$) across the phase transition suggests the involvement of higher-order coupling between the lattice and polarization in the KDP crystal. Furthermore, the bulk modulus ($B$) undergoes a sudden softening precisely at the transition, while the shear modulus ($G$) initially softens and subsequently stiffens across the transition.
\end{abstract}	
\end{frontmatter}
	
\section{Introduction}
Potassium dihydrogen phosphate (KH$_{2}$PO$_{4}$), commonly referred to as KDP, is a well-known dielectric material widely utilized in non-linear optical and electro-optical applications \cite{Salvo748,Yokotani1030, Yoreo113}. KDP undergoes a paraelectric to ferroelectric phase transition ($T_{C}\sim122$~K) accompanied by a tetragonal (space group $I\bar42d$) to orthorhombic ($Fdd2$) crystal symmetry change\cite{Reese504,Kobayashipssa63,Kobayashipssb293}. This phase transition has been a subject of interest for a long time, with some reports suggesting its proximity to a tricritical point\cite{Reese504,Reese510,Reese905,Schmidt839,Bastie337,Melo67,PETROVA26}. The phase transition in KDP has been attributed to changes in hydrogen bonding, leading to an abrupt jump in both the polarization and lattice parameters. This strong coupling between the lattice and polarization in this material has prompted extensive investigations into its elastic properties.The elastic properties have been probed using various acoustic techniques, including pulse-echo ultrasonic scattering, Brillouin scattering, and resonance frequency measurements\cite{Mason173, Brody179,Garland971,Hearmon120,Litov19,PETROVA26}. These techniques possess certain advantages and experimental constraints, such as the frequency range, sample size, and crystal orientation. As a result, they may yield different values for the elastic constants. Initial investigation using ultrasonic experiments on KDP unveiled anomalous behavior in the elastic constants $C_{33}$ and $C_{66}$ across the phase transition, which was attributed to strong piezoelectric coupling\cite{Garland971,Litov19,Mason173}. Furthermore, a recent study has also revealed anomalies in other longitudinal elastic constants, including  $C_{11}$, $C_{44}$, $C_{33}$ and $C_{66}$ across the transition\cite{PETROVA26}. However, the aspect of crystal symmetry change at the transition has not been taken into account in these ultrasonic experiments. In KDP crystal, the tetragonal crystal symmetry ($\bar42m$) consists of six elastic constants, while the orthorhombic symmetry ($2mm$) encompasses nine elastic constants. The evolution of the additional elastic constants in the low symmetry phase remain to be explored.

In this context, resonant ultrasound spectroscopy (RUS) has emerged as a compelling technique for the determination of elastic constants\cite{RUS1,RUS_may_rev,RUS2,RUS3,RUS4,RUS5,RUS6}. Unlike conventional ultrasonic methods, RUS enables the simultaneous measurement of all elastic constants within a single scan, providing a comprehensive characterization of a solid's elastic behavior. Moreover, the inverse of the mechanical quality factor Q$^{-1}$ obtained from RUS measurements offers insights into elastic relaxation phenomena\cite{RUS7}. Given that RUS is highly sensitive to crystal symmetry and provides a comprehensive set of elastic constants, it would be intriguing to investigate the thermal behavior of all the elastic constants of KDP crystal across the phase transition. To the best of our knowledge, RUS has solely been carried out on KDP at room temperature till date\cite{RUSKDP_RT}. 

In the present study, we conducted temperature-dependent RUS measurements on high-quality KDP single crystal within the temperature range of 100~K to 300~K. The results revealed a systematic elastic softening across the entire temperature range, except in the transition region where pronounced stiffness was observed near the transition temperature. The phase transition is effectively probed using the RUS data, as evidenced by the distinct patterns observed in the $f_{0}^{2}$ and $Q^{-1}$ plots. Additionally, anomalous behavior across in the all elastic constants further indicates the presence of a complex higher order piezoelectric coupling across the phase transition in KDP crystal.
\begin{figure}[htb]
	\begin{center}
		\includegraphics[width = \linewidth]{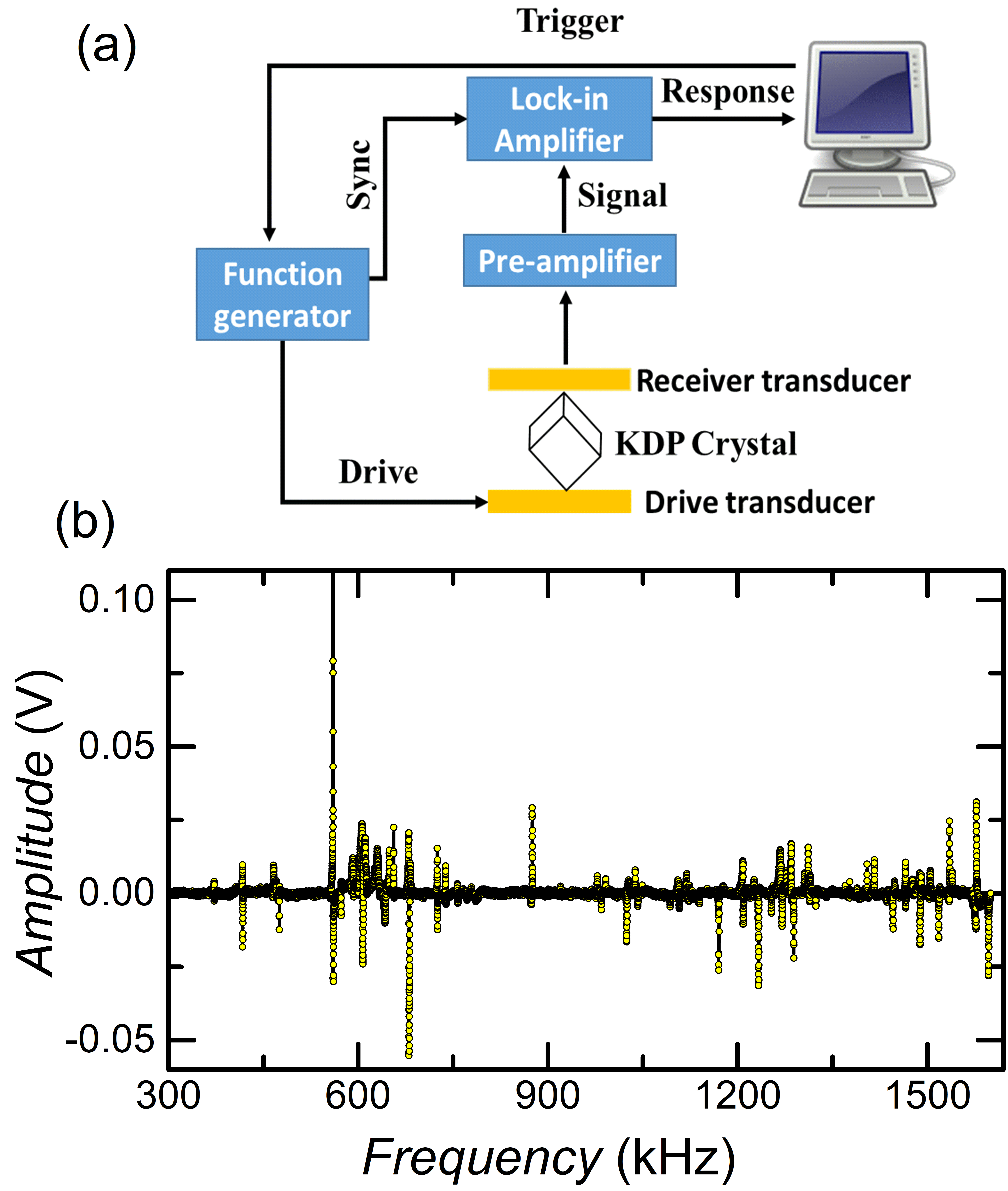}
	\end{center}
	\caption{(color online) \textbf{(a)}Schematic of the RUS measurement setup with the KDP crystal mounted between two transducers. \textbf{(b)} Amplitude of the measured RUS spectrum of the KDP crystal, where the peaks correspond to natural elastic resonances of KDP crystal.}
	\label{Fig1}
\end{figure}
\section{Experimental Details}
Resonant ultrasound spectroscopy (RUS) experiment is performed on high quality KDP single crystal. The single crystal of KDP is grown by slow evaporation of a supersaturated solution method. A rectangular parallelepiped sample is cut along the crystallographic axes and polished with dimensions of $2.3 \times3.1\times2.1$~mm$^{3}$ and mass density $2288$~kg/m$^{3}$. In the RUS experiment, the sample is delicately mounted between two Z-cut LiNbO$_{3}$ piezoelectric transducers. One of the transducers functions as the driver and other serves as the signal receiver or detector. To measure the resonances, the drive transducer is excited with a sinusoidal voltage signal of fixed frequency using a function generator (Tektronix AFG 1022) and the sample's response is detected by a Lock-In Amplifier (SRS SR844 RF) through a signal pre-amplifier. A typical schematic diagram is shown in Fig.\ref{Fig1}(a).  The RUS spectrum comprises a series of resonance frequencies that are intricately linked to the elastic constants, dimensions, density, and crystal symmetry of the sample. Resonance arises when the excitation frequency matches with a normal mode of the sample. To obtain a comprehensive set of elastic constants, a substantial number of resonances are indispensable.
	To capture the large number of resonance modes, excitation frequency is swept from 300~kHz to 1600~kHz and a measured RUS spectrum of KDP single crystal, depicting distinct resonances is shown in Fig.\ref{Fig1}(b). 
	\begin{figure}[t]
		\begin{center}
			\includegraphics[width = \linewidth]{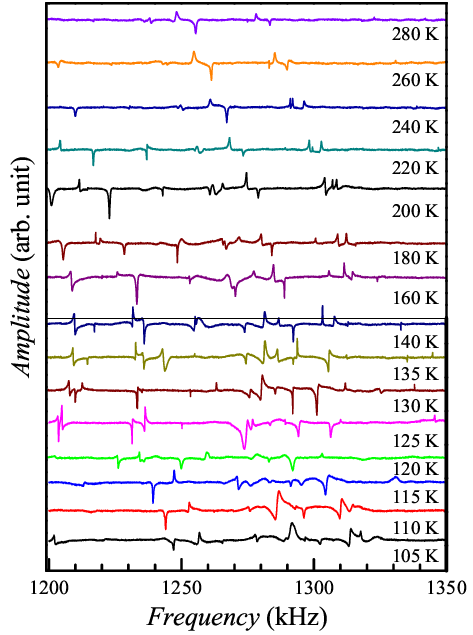}
		\end{center}
		\caption{(color online) Temperature-dependent resonant ultrasound spectra of KDP single crystal collected at different temperatures ranging from 280~K to 100~K, covering a frequency range of 1200~kHz to 1350~kHz. To enhance clarity, a y-axis offset is applied, and the measurement temperatures are labeled. Significant variations in the resonance peaks are clearly evident as the temperature changes.}
		\label{Fig2}
	\end{figure}
	\begin{figure*}[htb]
	\begin{center}
		\includegraphics[width = \linewidth]{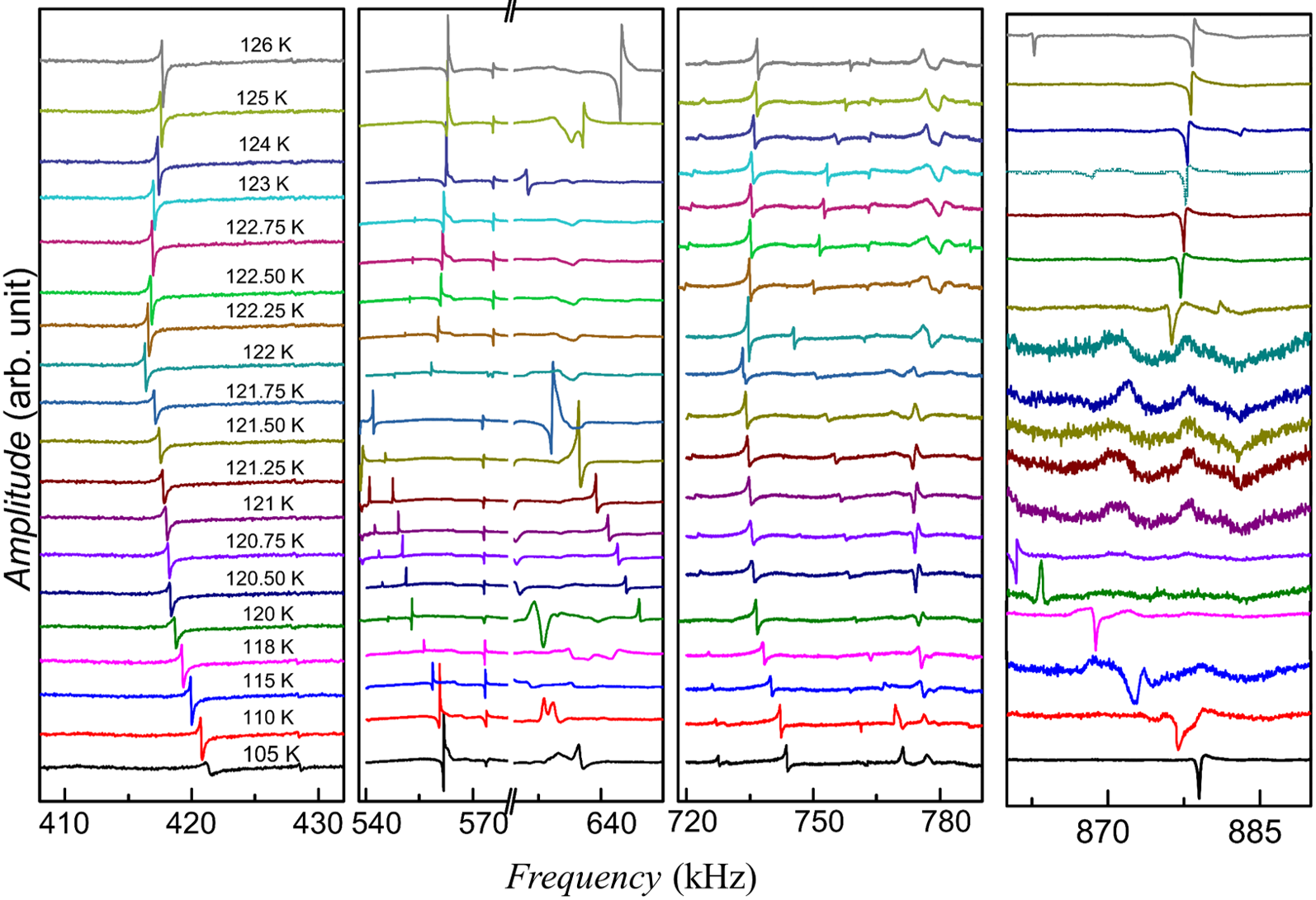}
	\end{center}
	\caption{(color online) Temperature-dependent RUS spectra collected within a narrow temperature interval from 126~K to 105~K, across the phase transition. The x-axis is divided into four selected segments, while each spectrum is offset on the y-axis to enhance visualization and clarity. The label indicates the corresponding data collection temperature. A distinct signature of the phase transition is evident in the alteration of resonance peaks.}
	\label{Fig3}
\end{figure*}
	In temperature-dependent measurements, the RUS setup is integrated with a high vacuum compatible variable temperature insert (VTI) that is immersed within a liquid nitrogen Dewar. Precise temperature monitoring and controlling are achieved through a Cernox sensor and a strip heater regulated by a Lakeshore 330 temperature controller. To control the instruments and acquire data seamlessly, we implemented a LabView program. The RUS spectra are acquired over a range of temperatures, spanning from 300~K to 100~K, with a reduced temperature interval of 0.25~K near the phase transition region. Prior to each frequency scan, temperature is well stabilized to within $\pm 10$~mK. The analysis of  RUS spectra involved employing of the Rayleigh-Ritz method\cite{RUS3,RUS5,Maynard2,RUS_orth_theory}. This method enabled us to determine the natural resonance frequencies by solving the elastic wave equation, taking into account the known elastic constants, dimensions, crystal symmetry, and density of the sample. Further, the least-squares method is utilized to refine the elastic constants by minimizing the difference between the observed and calculated resonances. Through this refinement process, the elastic constants are adjusted iteratively to achieve a better agreement between the experimental and theoretical resonance frequencies. 
	\begin{figure}[htb]
		\begin{center}
			\includegraphics[width = \linewidth]{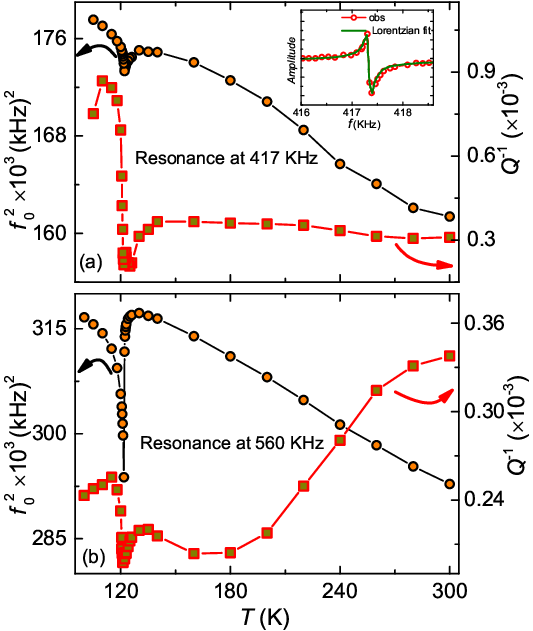}
		\end{center}
		\caption{(color online) Temperature dependent of $f_{0}^{2}$ (circle) and $Q^{-1}$ (square) for (a) resonance at 417~kHz and (b) 560~kHz. Insert of (a) contains the Lorentzian fit of corresponding resonance. Note: The error bar is smaller than the size of the symbol.}
		\label{Fig4}
	\end{figure}
\section{Results and Discussion}
Figure \ref{Fig2} illustrates the temperature-dependent RUS spectra, which are acquired within the frequency range of 1200~kHz to 1350~kHz. The spectra are presented with an offset for clarity. Observations reveal a noticeable shift towards lower frequency of the prominent resonances as the temperature increases, indicating elastic softening. However, a significant and abrupt alteration in the resonance pattern is evident across the phase transition temperature. Therefore, to probe the phase transition extensive RUS spectra are measured in the close interval of temperature. Figure \ref{Fig3} represents the RUS spectra, clearly indicating the occurrence of a ferroelectric phase transition. It is observed that certain resonance peaks vanish across the phase transition, signifying a change in the crystal symmetry. Furthermore, some resonances demonstrate elastic softening as the temperature increases, with a subsequent hardening observed at the transition. To elaborate on this, we have extracted the resonance frequency ($f_{0}$), mechanical quality factor ($Q = \Delta f/f_{0}$) by fitting the resonance peak with the Lorentzian function $A(f) = A_{0}\big(\frac{\frac{f}{f_{0}}\cos\Phi + (1-(\frac{f}{f_{0}})^{2})Q\sin\Phi}{(\frac{f}{f_{0}})^{2} + (1-(\frac{f^{2}}{f_{0}})^{2})Q^{2} }\big) + a_{1} + a_{2}f + a_{3}f^{2} + a_{4}f^{3}$, where $A_{0}$ is the peak amplitude, $Q$ is the quality factor, $\Phi$ is the phase and $f_{0}$ is the resonance frequency is of the peak and the last term represents a polynomial component incorporated to account the background signal in the fit. Figure\ref{Fig4}(a,b) shows the temperature dependent plot of $f_{0}^{2}$ and $Q^{-1}$ obtained from Lorentzian fit of the selected resonances around 417~kHz and 560~kHz. A Lorentzian fit is shown in insert of Fig.\ref{Fig4}(a). These resonances initially exhibit a gradual softening as the temperature increases. However, in the vicinity of the ferroelectric transition temperature, a reverse trend characterized by hardening becomes apparent. Following the transition, the softening resumes as the temperature continues to rise. These features can be distinctly observed as the anomalies in both $f_{0}^{2}$ and $Q^{-1}$ around the  122~K. The resonance at 560~kHz demonstrates a notable sensitivity to the ferroelectric to paraelectric phase transition, as it exhibits a pronounced attenuation at the transition temperature. The observed strong attenuation indicates the involvement of specific elastic modes associated with the transition.
\begin{table*}[hbt]
	\caption{Experimentally determined elastic constants ($C_{ij}$) for KH$_{2}$PO$_{4}$ (KDP) from RUS data at 300 K and 100 K}
	\label{Table1}
	\begin{tabular*}{\textwidth}{@{\extracolsep{\fill}}cccccccccc}
		\hline
		&&\multicolumn{6}{c}{Elastic constants (GPa)} \\
		System&$C_{11}$&$C_{12}$&$C_{13}$&$C_{22}$&$C_{23}$&$C_{33}$&$C_{44}$&$C_{55}$&$C_{66}$\\
		\hline
		Tetragonal (300 K)&79.32(1)&18.65(3)&9.10(1)&--&--&52.99(2)&13.71(5)&--&6.15(1)\\
		Orthorhombic (100 K)&60.56(2)&4.27(4)&1.58(2)&33.87(2)&14.42(4)&55.54(6)&14.84(3)&4.82(1)&8.77(2)\\
		\hline
	\end{tabular*}
\end{table*}
\begin{figure*}[hbt]
	\begin{center}
		\includegraphics[width = \linewidth]{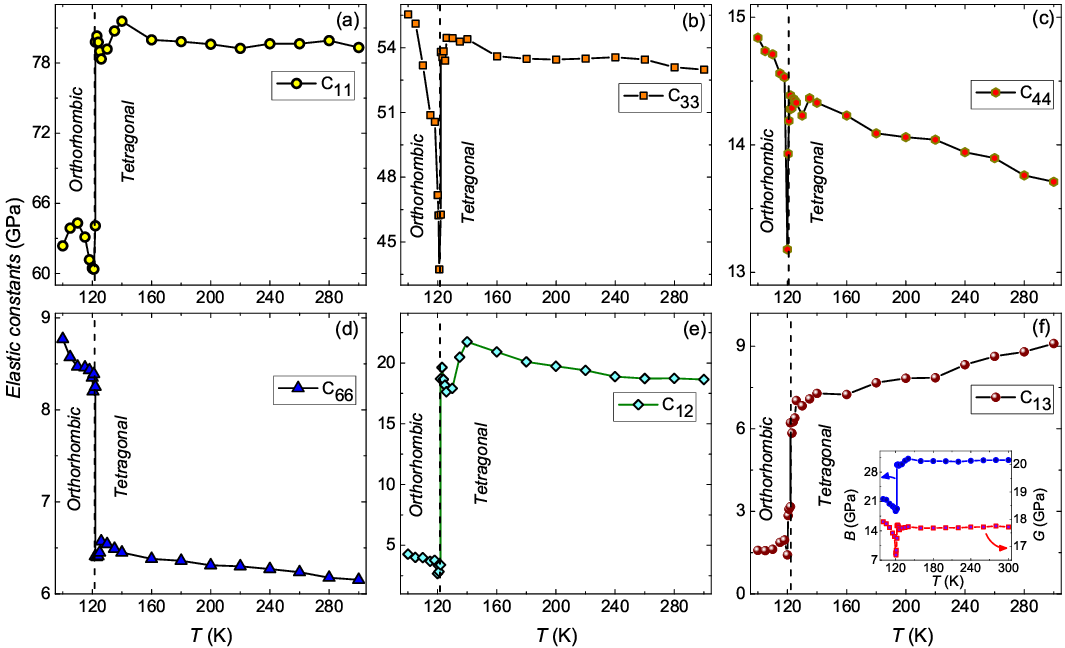}
	\end{center}
	\caption{(color online) Temperature dependence of elastic constants ($C_{11}$, $C_{33}$, $C_{44}$, $C_{66}$, $C_{12}$, and $C_{13}$) of KDP single crystal across the tetragonal to orthorhombic phase transition. Inset of (f): Temperature dependence of bulk modulus ($B$) and shear modulus ($G$).  Note: The error bar is smaller than the size of the symbol.}
	\label{Fig5}
\end{figure*}

The elastic constants of KDP are determined through an iterative process, wherein the previously reported values of the elastic constants are used as starting values.\cite{Litov19,HAUSS401,PETROVA26,Hearmon120} As mentioned earlier, KDP exhibits tetragonal crystal symmetry ($\bar{4}2m$) at room temperature, with six distinct elastic constants ($C_{11}$, $C_{33}$, $C_{44}$, $C_{66}$, $C_{12}$, $C_{13}$,). However, below 122~K, the crystal symmetry changes to orthorhombic ($2mm$), resulting in nine elastic constants ($C_{11}$, $C_{22}$, $C_{33}$, $C_{44}$, $C_{55}$, $C_{66}$, $C_{12}$, $C_{13}$, and $C_{23}$). In our analysis, the first 60 resonances are selected, and through an iterative procedure, we achieved a difference of $\le$0.3\% between the calculated and observed resonance frequencies (see the supplementary information). The determined elastic constants at 300~K (tetragonal phase) and 100~K (orthorhombic phase) are presented in Table~\ref{Table1}. The obtained values of the elastic constants at 300~K are reasonably close to the previously reported for the tetragonal symmetry\cite{Litov19,HAUSS401,PETROVA26}. The variation of all elastic constants with temperature is plotted in Fig. \ref{Fig5}. The observed linear behavior of the elastic constants in the paraelectric phase can be attributed to lattice anharmonicity. However, during the phase transition to the ferroelectric phase, the lattice anharmonicity becomes more pronounced. The presence of spontaneous polarization in the ferroelectric phase introduces additional interactions and structural changes, causing deviations from harmonic behavior. These deviations manifest as nonlinearities in the elastic response of the material. Across the phase transition, a discontinuity is observed in the elastic constants due to the crystal symmetry change. The elastic constant $C_{66}$ undergoes a significant stiffening of approximately $\sim$27\% during the paraelectric to the ferroelectric phase transition, while $C_{11}$ exhibits a softening of $\sim$24\% across the transition. Additionally, the off-diagonal constants $C_{12}$ and $C_{13}$ also experience substantial softening throughout this phase transition. In contrast, $C_{33}$ and $C_{44}$ exhibit a minimum value at the transition temperature. It is worth noting that although the absolute values of the elastic constants for tetragonal symmetry align with the reported values, there are noticeable differences in the thermal behavior of these constants across the transition\cite{Litov19,PETROVA26}. These disparities in the thermal variations of the elastic constants can be attributed to the explicit consideration of the change in crystal symmetry during the data analysis of the RUS, which distinguishes it from other ultrasonic methods. The anomalous behavior of elastic constants $C_{11}$, $C_{33}$, and $C_{66}$ at the phase transition could be a result of piezoelectric coupling between the lattice and polarization along the Z-direction. On the other hand, the anomalous behavior of the remaining elastic constants $C_{12}$, $C_{13}$, $C_{44}$, and $C_{55}$ suggests the involvement of higher-order piezoelectric coupling mechanisms. Furthermore, the bulk modulus ($B$) and shear modulus ($G$) of the KDP crystal are calculated from the Voigt-Reuss-Hill (VRH) approximations\cite{Voigt1928,Reuss49,Hill_1952}. In the VRH approximation, the bulk modulus ($B$) is determined as the Hill arithmetic average of the bulk Voigt ($B_{V}$) and Reuss ($B_{R}$) moduli. Similarly, the shear modulus ($G$) is obtained as the Hill arithmetic average of the shear Voigt ($G_{V}$) and Reuss ($G_{R}$) moduli. The Voigt and Reuss moduli represent the upper and lower limit of the actual effective moduli, respectively\cite{Voigt1928,Reuss49,Hill_1952,Wu054115}.The relationship of Voigt ($B_{V}$ and $G_{V}$) and Reuss ($B_{R}$ and $G_{R}$) moduli with elastic constants ($C_{ij}$) for tetragonal and orthorhombic phase can be found in Ref.~\cite{Tanju205201,Wu054115}. The obtained curves for both the bulk modulus ($B$) and shear modulus ($G$) are plotted in the inset of Fig.\ref{Fig5}(f). In this plot, it can be observed that the bulk modulus exhibits a softening behavior, indicating a decrease in stiffness, across the transition. On the other hand, the shear modulus exhibits a minimum value at the transition, suggesting a temporary reduction in resistance to shear deformation.
	
\section{Conclusion}
We have investigated the elastic properties of the KH$_{2}$PO$_{4}$ (KDP) single crystal across the paraelectric to ferroelectric phase transition using the resonant ultrasound spectroscopy (RUS) technique. The paraelectric and ferroelectric phase transition is clearly observed through anomalies in the temperature dependence of resonance frequencies and mechanical quality factor ($Q$). We have determined the full set of elastic constants for the both tetragonal and orthorhombic phases within the temperature range of 100-300~K. The thermal evolution of the elastic constants exhibits a pronounced change at the transition temperature ($T_{C}\sim122$~K). Specifically, the elastic constants $C_{11}$, $C_{12}$, and $C_{13}$ display a significant softening behavior during the transition, while $C_{66}$ experiences a stiffening effect. Additionally, $C_{33}$ and $C_{44}$ demonstrate a minimum value at the transition. The bulk modulus ($B$) exhibits a sudden softening at the transition, while the shear modulus ($G$) demonstrates an initial softening followed by stiffening across the transition.
	
\section{Acknowledgements}
We would like to acknowledge DST, India for financial support through Grant Number CRG/2019/004653.

\bibliographystyle{elsarticle-num}
\bibliography{reff_KDP2}

\providecommand{\noopsort}[1]{}\providecommand{\singleletter}[1]{#1}%
\begin{thebibliography}{10}
\expandafter\ifx\csname url\endcsname\relax
  \def\url#1{\texttt{#1}}\fi
\expandafter\ifx\csname urlprefix\endcsname\relax\def\urlprefix{URL }\fi
\expandafter\ifx\csname href\endcsname\relax
  \def\href#1#2{#2} \def\path#1{#1}\fi

\bibitem{Salvo748}
C.~Salvo, Solid-state light valve, IEEE Transactions on Electron Devices 18~(9)
  (1971) 748--755.
\newblock \href {https://doi.org/10.1109/T-ED.1971.17276}
  {\path{doi:10.1109/T-ED.1971.17276}}.

\bibitem{Yokotani1030}
A.~Yokotani, T.~Sasaki, K.~Yoshida, T.~Yamanaka, C.~Yamanaka, {Improvement of
  the bulk laser damage threshold of potassium dihydrogen phosphate crystals by
  ultraviolet irradiation}, App. Phys. Lett. 48~(16) (1986) 1030.
\newblock \href {https://doi.org/10.1063/1.96638} {\path{doi:10.1063/1.96638}}.

\bibitem{Yoreo113}
J.~J.~D. Yoreo, A.~K. Burnham, P.~K. Whitman, Developing {KH}$_{2}${PO}$_{4}$
  and {KD}$_{2}${PO}$_{4}$ crystals for the world's most power laser, Int.
  Mater. Rev. 47~(3) (2002) 113--152.
\newblock \href {https://doi.org/10.1179/095066001225001085}
  {\path{doi:10.1179/095066001225001085}}.

\bibitem{Reese504}
W.~Reese, L.~F. May, Studies of phase transitions in order-disorder
  ferroelectrics. {II}. calorimetric investigations of {KD}$_2${PO}$_{4}$,
  Phys. Rev. 167 (1968) 504--510.
\newblock \href {https://doi.org/10.1103/PhysRev.167.504}
  {\path{doi:10.1103/PhysRev.167.504}}.

\bibitem{Kobayashipssa63}
J.~Kobayashi, Y.~Uesu, I.~Mizutani, Y.~Enomoto, X-ray study on thermal
  expansion of ferroelectric {KH}$_2${PO}$_{4}$, Phys. Status Solidi a 3~(1)
  (1970) 63.
\newblock \href {https://doi.org/https://doi.org/10.1002/pssa.19700030108}
  {\path{doi:https://doi.org/10.1002/pssa.19700030108}}.

\bibitem{Kobayashipssb293}
J.~Kobayashi, Y.~Uesu, Y.~Enomoto, X-ray dilatometric study on ferroelectric
  phase transition of {KH}$_2${PO}$_{4}$ {I}. the order of transition, Phys.
  Status Solidi b 45~(1) (1971) 293--304.
\newblock \href {https://doi.org/https://doi.org/10.1002/pssb.2220450133}
  {\path{doi:https://doi.org/10.1002/pssb.2220450133}}.

\bibitem{Reese510}
W.~Reese, L.~F. May, Critical phenomena in order-disorder ferroelectrics. i.
  calorimetric studies of {KH}$_2${PO}$_{4}$, Phys. Rev. 162 (1967) 510--518.
\newblock \href {https://doi.org/10.1103/PhysRev.162.510}
  {\path{doi:10.1103/PhysRev.162.510}}.

\bibitem{Reese905}
W.~Reese, Studies of phase transitions in order-disorder ferroelectrics. {III}.
  the phase transition in {KH}$_2${PO}$_{4}$ and a comparison with
  {KD}$_2${PO}$_{4}$, Phys. Rev. 181 (1969) 905--919.
\newblock \href {https://doi.org/10.1103/PhysRev.181.905}
  {\path{doi:10.1103/PhysRev.181.905}}.

\bibitem{Schmidt839}
V.~H. Schmidt, A.~B. Western, A.~G. Baker, Tricritical point in
  {KH}$_{2}${PO}$_{4}$, Phys. Rev. Lett. 37 (1976) 839.
\newblock \href {https://doi.org/10.1103/PhysRevLett.37.839}
  {\path{doi:10.1103/PhysRevLett.37.839}}.

\bibitem{Bastie337}
P.~Bastie, M.~Vallade, C.~Vettier, C.~M.~E. Zeyen, Study of the tricritical
  point in {KH}$_{2}${PO}$_{4}$ by $\ensuremath{\gamma}$-ray and neutron
  diffractometry, Phys. Rev. Lett. 40 (1978) 337.
\newblock \href {https://doi.org/10.1103/PhysRevLett.40.337}
  {\path{doi:10.1103/PhysRevLett.40.337}}.

\bibitem{Melo67}
F.~E.~A. Melo, S.~G.~C. Moreira, A.~S. Chaves, I.~Guedes, P.~T.~C. Freire,
  J.~Mendes-filho, Phase diagram uniaxial pressure—temperature of {KDP},
  Ferroelectrics 233~(1) (1999) 67.
\newblock \href {https://doi.org/10.1080/00150199908016997}
  {\path{doi:10.1080/00150199908016997}}.

\bibitem{PETROVA26}
A.~Petrova, S.~Stishov, Elastic properties of {KH}$_{2}${PO}$_{4}$ at the
  ferroelectric phase transition, Solid State Communications 171 (2013) 26.
\newblock \href {https://doi.org/https://doi.org/10.1016/j.ssc.2013.07.029}
  {\path{doi:https://doi.org/10.1016/j.ssc.2013.07.029}}.

\bibitem{Mason173}
W.~P. Mason, The elastic, piezoelectric, and dielectric constants of potassium
  dihydrogen phosphate and ammonium dihydrogen phosphate, Phys. Rev. 69 (1946)
  173.
\newblock \href {https://doi.org/10.1103/PhysRev.69.173}
  {\path{doi:10.1103/PhysRev.69.173}}.

\bibitem{Brody179}
E.~M. Brody, H.~Z. Cummins, Brillouin-scattering study of the elastic anomaly
  in ferroelectric {KH}$_2${PO}$_{4}$, Phys. Rev. B 9 (1974) 179.
\newblock \href {https://doi.org/10.1103/PhysRevB.9.179}
  {\path{doi:10.1103/PhysRevB.9.179}}.

\bibitem{Garland971}
C.~W. Garland, D.~B. Novotny, Ultrasonic velocity and attenuation in
  {KH}$_2${PO}$_{4}$, Phys. Rev. 177 (1969) 971.
\newblock \href {https://doi.org/10.1103/PhysRev.177.971}
  {\path{doi:10.1103/PhysRev.177.971}}.

\bibitem{Hearmon120}
R.~F.~S. Hearmon, \href{https://dx.doi.org/10.1088/0508-3443/3/4/303}{The
  elastic constants of piezoelectric crystals}, Brit. J. Appl. Phys. 3~(4)
  (1952) 120.
\newblock \href {https://doi.org/10.1088/0508-3443/3/4/303}
  {\path{doi:10.1088/0508-3443/3/4/303}}.
\newline\urlprefix\url{https://dx.doi.org/10.1088/0508-3443/3/4/303}

\bibitem{Litov19}
E.~Litov, C.~W. Garland, Ultrasonic experiments in kdp-type crystals,
  Ferroelectrics 72~(1) (1987) 19.
\newblock \href {https://doi.org/10.1080/00150198708017936}
  {\path{doi:10.1080/00150198708017936}}.

\bibitem{RUS1}
J.~Maynard, {Resonant Ultrasound Spectroscopy}, Physics Today 49~(1) (1996)
  26--31.
\newblock \href {https://doi.org/10.1063/1.881483}
  {\path{doi:10.1063/1.881483}}.

\bibitem{RUS_may_rev}
J.~D. Maynard, {Resonance spectroscopy for solids with layers of different
  materials}, JASA Express Letters 2~(12) (12 2022).
\newblock \href {https://doi.org/10.1121/10.0015316}
  {\path{doi:10.1121/10.0015316}}.

\bibitem{RUS2}
A.~Migliori, J.~Sarrao, W.~M. Visscher, T.~Bell, M.~Lei, Z.~Fisk, R.~Leisure,
  Resonant ultrasound spectroscopic techniques for measurement of the elastic
  moduli of solids, Phys. B: Condens. Matter 183~(1) (1993) 1--24.
\newblock \href {https://doi.org/https://doi.org/10.1016/0921-4526(93)90048-B}
  {\path{doi:https://doi.org/10.1016/0921-4526(93)90048-B}}.

\bibitem{RUS3}
R.~G. Leisure, F.~A. Willis, Resonant ultrasound spectroscopy, J. Phys.
  Condens. Matter 9~(28) (1997) 6001.
\newblock \href {https://doi.org/10.1088/0953-8984/9/28/002}
  {\path{doi:10.1088/0953-8984/9/28/002}}.

\bibitem{RUS4}
B.~J. Zadler, J.~H.~L. Le~Rousseau, J.~A. Scales, M.~L. Smith, {Resonant
  Ultrasound Spectroscopy: theory and application}, Geophys. J. Int. 156~(1)
  (2004) 154--169.
\newblock \href {https://doi.org/10.1111/j.1365-246X.2004.02093.x}
  {\path{doi:10.1111/j.1365-246X.2004.02093.x}}.

\bibitem{RUS5}
A.~Migliori, J.~D. Maynard, {Implementation of a modern resonant ultrasound
  spectroscopy system for the measurement of the elastic moduli of small solid
  specimens}, Rev. Sci. Instrum. 76~(12) (12 2005).
\newblock \href {https://doi.org/10.1063/1.2140494}
  {\path{doi:10.1063/1.2140494}}.

\bibitem{RUS6}
F.~F. Balakirev, S.~M. Ennaceur, R.~J. Migliori, B.~Maiorov, A.~Migliori,
  {Resonant ultrasound spectroscopy: The essential toolbox}, Rev. Sci. Instrum.
  90~(12) (12 2019).
\newblock \href {https://doi.org/10.1063/1.5123165}
  {\path{doi:10.1063/1.5123165}}.

\bibitem{RUS7}
D.~Yang, G.~I. Lampronti, C.~R.~S. Haines, M.~A. Carpenter, Magnetoelastic
  coupling behavior at the ferromagnetic transition in the partially disordered
  double perovskite {La}$_{2}${NiMnO}$_{6}$, Phys. Rev. B 100 (2019) 014304.
\newblock \href {https://doi.org/10.1103/PhysRevB.100.014304}
  {\path{doi:10.1103/PhysRevB.100.014304}}.

\bibitem{RUSKDP_RT}
T.~Liu, G.~Pei, Z.~Zhang, H.~Wang, {Surface distortion prediction method of
  {KDP} frequency converters}, in: Y.~Huang (Ed.), Tenth International
  Conference on Information Optics and Photonics, Vol. 10964, International
  Society for Optics and Photonics, SPIE, 2018, p. 109645J.
\newblock \href {https://doi.org/10.1117/12.2506374}
  {\path{doi:10.1117/12.2506374}}.

\bibitem{Maynard2}
G.~Liu, J.~D. Maynard, {Measuring elastic constants of arbitrarily shaped
  samples using resonant ultrasound spectroscopy}, The Journal of the
  Acoustical Society of America 131~(3) (2012) 2068.
\newblock \href {https://doi.org/10.1121/1.3677259}
  {\path{doi:10.1121/1.3677259}}.

\bibitem{RUS_orth_theory}
I.~OHNO, Free vibration of a rectangular parallelepiped crystal and its
  application to determination of elastic constants of orthorhombic crystals,
  Journal of Physics of the Earth 24~(4) (1976) 355--379.
\newblock \href {https://doi.org/10.4294/jpe1952.24.355}
  {\path{doi:10.4294/jpe1952.24.355}}.

\bibitem{HAUSS401}
S.~HAUSS{Ü}HL, \href{https://doi.org/10.1524/zkri.1964.120.16.401}{Elastische
  und thermoelastische eigenschaften von {KH}$_{2}${PO}$_{4}$,
  {KH}$_{2}${A}s{O}$_{4}$, {NH}$_{4}${H}$_{2}${PO}$_{4}$,
  {NH}$_{4}${H}$_{2}${A}s{O}$_4$ und {R}b{H}$_{2}${PO}$_{4}$}, Z. Kristallogr.
  Cryst. Mater. 120~(1-6) (1964) 401.
\newblock \href {https://doi.org/doi:10.1524/zkri.1964.120.16.401}
  {\path{doi:doi:10.1524/zkri.1964.120.16.401}}.
\newline\urlprefix\url{https://doi.org/10.1524/zkri.1964.120.16.401}

\bibitem{Voigt1928}
W.~Voigt, Lehrbuch der kristallphysik: Teubner-leipzig (1928).

\bibitem{Reuss49}
A.~Reuss, Berechnung der {F}lie{$\beta$}grenze von {M}ischkristallen auf
  {G}rund der {P}lastizit{ä}tsbedingung f{ü}r einkristalle ., Z. Angew. Math.
  Mech. 9~(1) (1929) 49--58.
\newblock \href {https://doi.org/https://doi.org/10.1002/zamm.19290090104}
  {\path{doi:https://doi.org/10.1002/zamm.19290090104}}.

\bibitem{Hill_1952}
R.~Hill, \href{https://dx.doi.org/10.1088/0370-1298/65/5/307}{The elastic
  behaviour of a crystalline aggregate}, Proc. Phys. Soc. A 65~(5) (1952) 349.
\newblock \href {https://doi.org/10.1088/0370-1298/65/5/307}
  {\path{doi:10.1088/0370-1298/65/5/307}}.
\newline\urlprefix\url{https://dx.doi.org/10.1088/0370-1298/65/5/307}

\bibitem{Wu054115}
Z.-j. Wu, E.-j. Zhao, H.-p. Xiang, X.-f. Hao, X.-j. Liu, J.~Meng,
  \href{https://link.aps.org/doi/10.1103/PhysRevB.76.054115}{Crystal structures
  and elastic properties of superhard {I}r{N}$_{2}$ and {I}r{N}$_{3}$ from
  first principles}, Phys. Rev. B 76 (2007) 054115.
\newblock \href {https://doi.org/10.1103/PhysRevB.76.054115}
  {\path{doi:10.1103/PhysRevB.76.054115}}.
\newline\urlprefix\url{https://link.aps.org/doi/10.1103/PhysRevB.76.054115}

\bibitem{Tanju205201}
T.~G\"urel, C.~Sevik, T.~\ifmmode \mbox{\c{C}}\else \c{C}\fi{}a\ifmmode
  \breve{g}\else \u{g}\fi{}\ifmmode \imath \else~\i \fi{}n,
  \href{https://link.aps.org/doi/10.1103/PhysRevB.84.205201}{Characterization
  of vibrational and mechanical properties of quaternary compounds
  {Cu}${}_{2}${ZnSn}${S}_{4}$ and {C}u$_{2}${ZnSnSe}$_{4}$ in kesterite and
  stannite structures}, Phys. Rev. B 84 (2011) 205201.
\newblock \href {https://doi.org/10.1103/PhysRevB.84.205201}
  {\path{doi:10.1103/PhysRevB.84.205201}}.
\newline\urlprefix\url{https://link.aps.org/doi/10.1103/PhysRevB.84.205201}

\end{thebibliography}

\end{document}